\documentclass[aps,preprint,preprintnumbers,amsmath,amssymb,amsfonts,groupedaddress,superscriptaddress,showpacs,showkeys,floatfix]{revtex4-1}

\usepackage{graphicx}
\usepackage{subfigure}
\usepackage{bm}
\usepackage{amsmath}
\usepackage{color}
\usepackage{array}
\usepackage{CJK}

\newcommand{\bmath}{\begin{mathletters}}
\newcommand{\emath}{\end{mathletters}}
\newcommand{\be}{\begin{eqnarray}}
\newcommand{\ee}{\end{eqnarray}}
\newcommand{\ba}{\begin{array}}
\newcommand{\ea}{\end{array}}

\newcommand{\pr}{\prime}

\begin{document}
\title{A simultaneous feedback and feed-forward control and its application
to realize a random walk on the Bloch sphere in a superconducting Xmon-qubit system}

\author{Liang Xiang}
 \affiliation{Zhejiang Province Key Laboratory of Quantum Technology and Device, Department of Physics, Zhejiang University, Hangzhou, 310027, China}
\author{Zhiwen Zong}
 \affiliation{Zhejiang Province Key Laboratory of Quantum Technology and Device, Department of Physics, Zhejiang University, Hangzhou, 310027, China}
 \author{Zhenhai Sun}
 \affiliation{Zhejiang Province Key Laboratory of Quantum Technology and Device, Department of Physics, Zhejiang University, Hangzhou, 310027, China}
  \author{Ze Zhan}
 \affiliation{Zhejiang Province Key Laboratory of Quantum Technology and Device, Department of Physics, Zhejiang University, Hangzhou, 310027, China}
   \author{Ying Fei}
 \affiliation{Zhejiang Province Key Laboratory of Quantum Technology and Device, Department of Physics, Zhejiang University, Hangzhou, 310027, China}
  \author{Zhangjingzi Dong}
 \affiliation{Zhejiang Province Key Laboratory of Quantum Technology and Device, Department of Physics, Zhejiang University, Hangzhou, 310027, China}
 \author{Chongxin Run}
 \affiliation{Zhejiang Province Key Laboratory of Quantum Technology and Device, Department of Physics, Zhejiang University, Hangzhou, 310027, China}
 \author{Zhilong Jia}
 \affiliation{Key Laboratory of Quantum Information, University of Science and Technology of China, Hefei, 230026, China}
 \author{Peng Duan}
 \affiliation{Key Laboratory of Quantum Information, University of Science and Technology of China, Hefei, 230026, China}
 \author{Jianlan Wu }
 \email{jianlanwu@zju.edu.cn}
 \affiliation{Zhejiang Province Key Laboratory of Quantum Technology and Device, Department of Physics, Zhejiang University, Hangzhou, 310027, China}
 \author{Yi Yin}
 \email{yiyin@zju.edu.cn}
 \affiliation{Zhejiang Province Key Laboratory of Quantum Technology and Device, Department of Physics, Zhejiang University, Hangzhou, 310027, China}
 \author{Guoping Guo}
 \email{gpguo@ustc.edu.cn}
 \affiliation{Key Laboratory of Quantum Information, University of Science and Technology of China, Hefei, 230026, China}
 \affiliation{Origin Quantum Computing, Hefei, 230026, China}

\begin{abstract}
Measurement-based feedback control is central in quantum computing and precise quantum control.
Here we realize a fast and flexible field-programmable-gate-array-based feedback control in a
superconducting Xmon qubit system. The latency of room-temperature electronics is custom optimized
to be as short as 140 ns. Projective measurement of a signal qubit produces a feedback
tag to actuate a conditional pulse gate to the qubit. In a feed-forward process,
the measurement-based feedback tag is brought to a different target qubit for a conditional
control. In a two-qubit experiment, the feedback and feed-forward controls are
simultaneously actuated in consecutive steps. A quantum number is then generated by the
signal qubit, and a random walk of the target qubit is correspondingly triggered
and realized on the Bloch sphere. Our experiment provides a conceptually simple and intuitive
benchmark for the feedback control in a multi-qubit system. The feedback system
can be further scaled up for more complex feedback control experiments.

\end{abstract}

\maketitle


\section{Introduction}

Quantum feedback is an important element to realize fault-tolerant quantum computation
in a complex multi-qubit system~\cite{ChuangBook}. Quantum feedback is defined as a
conditional action back to the quantum system, based on the result of quantum state
measurement of qubits in the original system~\cite{SarovarPRA05, YamamotoPRX14}.
Quantum error correction depends on repeated measurements of qubit state and the feedback
control to correct the error in a redundant quantum system~\cite{ChuangBook}.
The superconducting qubit system is a promising platform to develop the quantum
feedback and quantum error correction~\cite{KrantzAPR19}. Although an autonomous
or coherent feedback can be realized without any external logical decision hardware
~\cite{GeerlingsPRL13,ShankarNat13, KellyNat15,LiuPRX16,AndersenArxiv19}, the measurement-based
feedback control is a natural choice to provide feedback to the quantum system
~\cite{VijayNat12,RistePRL12,RisteNat13,SteffenNat13,CampagnePRX13,LangePRL14,OfekNat16,RyanRSI17,
MasuyamaNat18,SalathPRAppl18,HuNat19,AndersenNpjQi19}, with the controller
itself a classical instrument.

In different measurment-based feedback control protocols, both analog feedback and
digital feedback systems have been developed~\cite{VijayNat12, RistePRL12, RisteNat13,
SteffenNat13, CampagnePRX13, LangePRL14,OfekNat16,RyanRSI17,MasuyamaNat18,SalathPRAppl18,HuNat19,AndersenNpjQi19}.
The analog feedback is often based on partial measurements and acts to
the qubit with continuous parameter~\cite{VijayNat12, LangePRL14}.
The digital feedback based on projective measurement is very flexible and can be
directly applied to multi-qubit protocols. A relative fast digital feedback
control~\cite{RyanRSI17, SalathPRAppl18, HuNat19, AndersenNpjQi19}
is promising for the future complex applications~\cite{BarendsNat14,FuMICRO17,GambettaNPJ17,AruteNat19}.
For this digital feedback control, different combinations have been made to be
compatible with the requirement of experiment. For example, a commercial
field-programmable-gate-array (FPGA) based card (Nallatech BenADDA-V4) is
combined with an arbitrary waveform generator (AWG) to control the qubit
system~\cite{SalathPRAppl18}. In quantum error correction
experiments~\cite{OfekNat16,HuNat19}, commercial FPGA boards (Innovative X6-1000M)
are applied for both the signal detection and the waveform generation.

For maximum flexibility, we custom design and make a fast feedback control
system in a multi-qubit framework, following a previous multi-board
architecture~\cite{chenyuAPL12}. We choose a relatively advanced FPGA
chip (Xilinx Kintex-7) for all the boards, and design each board
either as a measure board or as a control board. The separation
of measure and control boards helps to make full use of hardware
resources. The latency of each board and the network latency of
multi-boards are optimized in the design (see Appendix C). With measurement
and analysis, the feedback latency of room-temperature electronics
is estimated to be as short as 140 ns. Although in our current status
the delay of feedback loop is mainly limited by the relatively long
measurement pulse, the short latency of electronics will speed up
the whole process when a Purcell filter is included for a fast
measurement~\cite{ReedAPL10, JeffreyPRL14,KrantzAPR19}.

In the custom FPGA-based multi-boards, the board programming model
is also critical for a complex multi-qubit feedback
experiment~\cite{RyanRSI17,FuMICRO17}. We define the measure/control
instructions set architecture (ISA) for the measure/control
board (see Appendix D). Each control instruction directs the board
to read its waveform from memory and stream data to the D/A converters.
Each measure instruction directs the board receiving data
from the A/D converters and mix it with the reference waveform
from memory. The execution of multiple instructions is
synchronous in multiple boards and inter-coupled through the
feedback network.

Different benchmarks of the feedback control have been designed
and presented previously~\cite{VijayNat12, RistePRL12,
CampagnePRX13, SteffenNat13, AndersenNpjQi19}. The feedback control
in our system is all based on a high-fidelity projective measurement of
the superconducting qubit. The function of feedback is initially
proved by a reset experiment, in which a single qubit is
reset to the ground state with a feedback-reset-gate~\cite{RistePRL12}.
The qubit is further prepared by the feedback to an arbitrary known
state on the Bloch sphere~\cite{CampagnePRX13}, with the superposition
state $(|0\rangle+|1\rangle)/\sqrt{2}$ as an example. For both
single-qubit experiments, the feedback control can be applied
consecutively for multiple times, which enables a high-fidelity
qubit reset and a repeated preparation of a known qubit state.

The feedback control in a multi-qubit system is often
applied to measure the parity of entangled qubits to create
tag for following conditional gates~\cite{RisteNat13, AndersenNpjQi19}.
We design a simple and intuitive two-qubit experiment by choosing
a signal qubit and a target qubit. The signal qubit is measured to
produce a tag, which can be either sent to the signal qubit to
create a feedback control or sent to the target qubit to create
a feed-forward control. When the feedback and feed-forward control
simultaneously take effect, specified functions can be realized.
In a random walk experiment, the signal qubit is repeatedly prepared at
the state $(|0\rangle+|1\rangle)/\sqrt{2}$ and creates a series of
tags by the feedback, functioning as a random number generator (0 or 1).
Depending on the tag of each step, the target qubit is actuated
to rotate clockwise or anticlockwise with a designated angle.
The simultaneous feed-forward control thus leads to
a random walk of the target qubit on the Bloch sphere. We present
measured results for a random walk with one step, two steps and
three steps. Numerical simulation with master equation is applied
to analyze the rotation angle of target qubit in each path of the
random walk (see Appendix E). The simulation result indicates that the qubit
relaxation and dephasing induce the experimentally observed deviation
of the rotation angle away from ideal values.

The realization of qubit random walk experiment proves that our
multi-board control system is very appropriate for such complex quantum feedback task.
The custom system integrates both signal detection and the
waveform generation, enabling the optimization of hardware resources
and integrated feedback latency. With the same hardware and ISA, the system
can be further scaled up to support feedback/feed-forward experiments
with multiple signal qubits and multiple target qubits.

\section{Experimental System Including the Feedback Control}

A superconducting Xmon qubit~\cite{BarendsPRL13} sample was fabricated on a silicon substrate.
The substrate was initially immersed in buffered hydrofluoric
acid to remove native oxide. Afterwards it was loaded into an electron
beam evaporator, and deposited with an aluminum film. The resonator and
control-line structure in the sample were patterned in a stepper and
dry-etched with BCl$_3$/Cl$_2$ in an inductively coupled plasma etcher.
The Josephson junction structure was patterned by an electron beam
lithography and constructed by the aluminum double-angle evaporation.
A `bandage' electrical contact was also included in the junction
fabrication~\cite{DunsworthAPL17}.

\begin{figure}[htp]
\centering
\includegraphics[width=0.9\columnwidth]{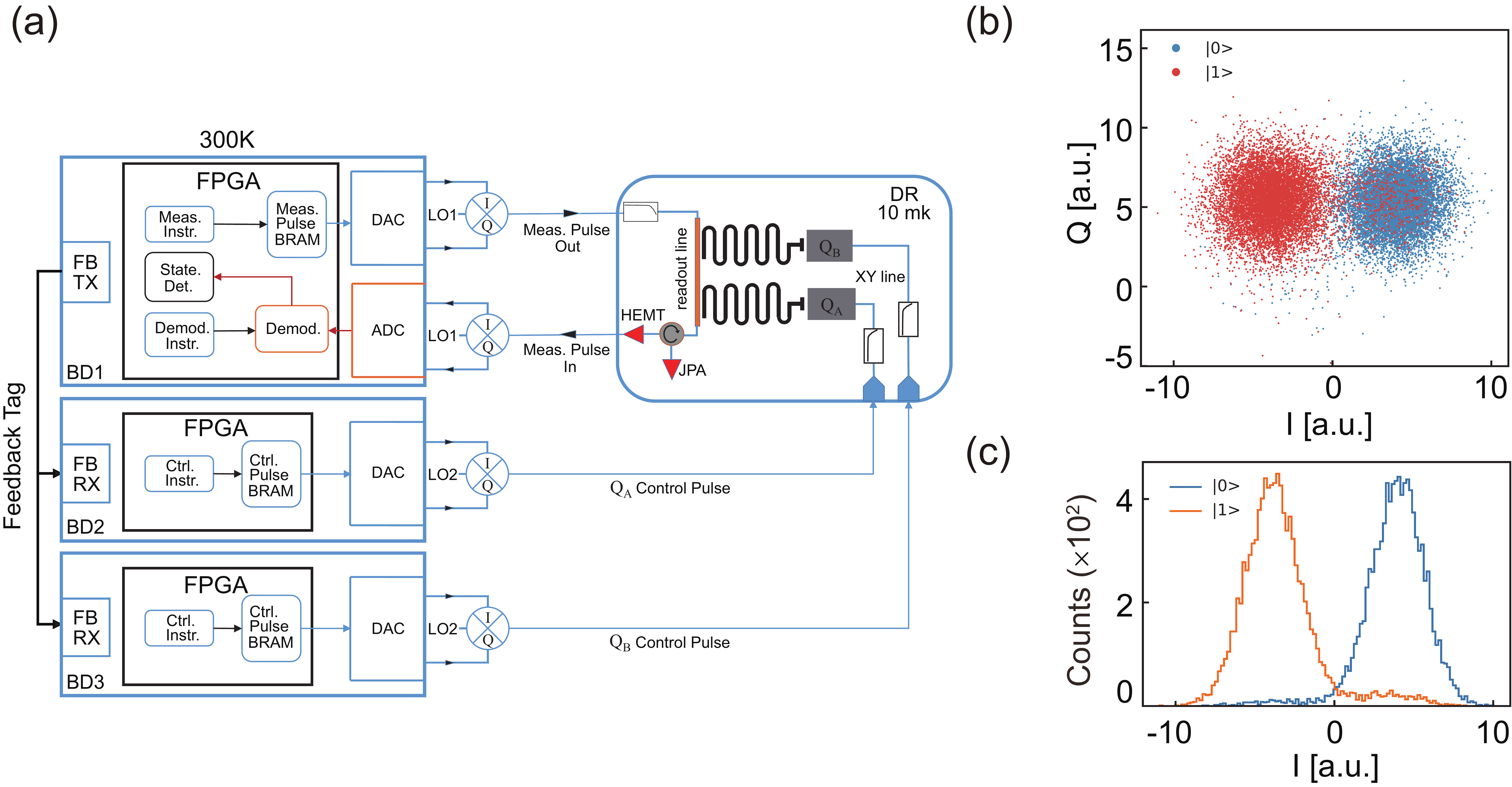}
\caption{
(a) Schematic diagram of the experimental system. Three customized FPGA-based boards are main control electronics of
the feedback loop. The board BD1 contains 1 FPGA mother board and 2 DAC/ADC daughter cards. The FPGA mother board reads
experimental instructions, fetches the measure pulse preloaded at the block random access memory (BRAM), and streams it
to the DAC card. The measure pulse at base-band ($<500$ MHz) is up-converted to the carrier frequency using an
analog $IQ$ mixer and a microwave signal (LO1). Another $IQ$ mixer down-converts the returned measure pulse to the
base-band, and the ADC card digitize the $IQ$ signals. The digital $IQ$ signals are further processed by the
demodulation logic and state determination logic. The feedback transmit (TX) logic
encapsulates the measurement result (0 or 1) to create a feedback tag, which is broadcasted to other boards. The boards
BD2 and BD3 receive the feedback tag and accordingly generate the conditional qubit control pulse.
(b) With Q$_\mathrm{B}$ initialized in the ground state $|0\rangle$ (blue) or excited state $|1\rangle$ (red), the measured
results are displayed in the $IQ$ plane with blue or red points, respectively.
(c) Statistical histogram of (b) on the $I$ axis, after data points integrated with respect to the $Q$ axis.
}
\label{fig_n01}
\end{figure}

A schematic diagram of the experimental setup is shown in Fig.~\ref{fig_n01}(a).
The qubit chip is mounted in a dilution refrigerator (DR) at a base
temperature of 10 mK. Through some cryogenic filters and
amplifiers, input and output lines of the qubit system are
connected to corresponding room-temperature electronics.
We mainly list three FPGA-based boards, which are key elements
in the feedback loop to measure, control and provide
feedback signal to the superconducting qubit system.

Two Xmon qubits Q$_\mathrm{A}$ and Q$_\mathrm{B}$ are
utilized in the following feedback experiment.
They are physically separated by three qubits in a linear
array of capacitively-coupled qubits, with the similar
qubit and chip described before~\cite{ZZXNJP2018, WangNJP2018, WangPRAppl2019}.
Two Xmon qubits are both biased at the sweet point~\cite{KochPRA07}
with a fixed operation frequency of $\omega_\mathrm{qA}/2\pi=5.050$ GHz
and $\omega_\mathrm{qB}/2\pi=5.079$ GHz, respectively.
The energy relaxation time $T_1$ are 16 $\mu$s and 19 $\mu$s,
and the pure dephasing time $T_2^*$ are 20 $\mu$s and 33 $\mu$s
for Q$_\mathrm{A}$ and Q$_\mathrm{B}$, respectively.
The qubit state is readout by a dispersive method
through a coupled readout resonator. The resonator bare
frequencies are $\omega_\mathrm{rA}/2\pi=6.521$ GHz
and $\omega_\mathrm{rB}/2\pi=6.438$ GHz for Q$_\mathrm{A}$
and Q$_\mathrm{B}$, respectively. In the dispersive readout,
a shaped readout pulse is sent through the readout line,
and encode a qubit-state-dependent dispersive shift
of the readout resonator. The readout signal is then
amplified by a low-temperature Josephson parametric
amplifier (JPA)~\cite{RoyAPL15} and a high electron
mobility transistor (HEMT). The amplified signal finally
goes into room temperature electronics for a data
collection and analysis.

The first FPGA board (BD1) is specially designed for
the function of qubit-state readout, called a measure
board. With the FPGA algorithm, two digital-to-analog-converters
(DACs) output shaped pulses with a carrier frequency smaller than
the DAC Nyquist frequency of 500 MHz. They are sent out to
an $IQ$ mixer as two quadratures of voltage, $I(t)$ and $Q(t)$.
A microwave source provides a local oscillator (LO) signal
($f=6.475$ GHz) for the $IQ$ mixer, which is modulated by
the $IQ$ quadratures to generate a readout signal with
variable frequency, phase and amplitude~\cite{KrantzAPR19}.
For the detection of returned signal, another $IQ$ mixer
mixes down the signal to two $IQ$ quadratures, which are
further digitized by two analog-to-digital-converters
(ADCs) in BD1. With a demodulation processing (see Appendix C),
the readout result is represented as data points in a
two-dimensional plane spanned by $I$ and $Q$.

In our experiment, the readout pulse has a quick initial
overshoot and a following sustain part~\cite{JeffreyPRL14,MalletNat09}.
The readout pulse is applied for a projective measurement,
in which the qubit is projected to the ground $|0\rangle$
or excited $|1\rangle$ state with probabilities
determined by the final qubit state before measurement.
With a 800 ns long readout pulse and a repetition of the
same measurement for many times ($C_\mathrm{tot}=1\times10^4$),
a typical distribution of readout result for Q$_\mathrm{B}$
is shown in the $IQ$ plane in Fig.~\ref{fig_n01}(b). The
blue and red dots represent result for the ground $|0\rangle$
and excited $|1\rangle$ state, respectively. They are observed
to be two separated humps of data points. The distribution
of each hump can be fitted with a Gaussian function, from
which a center of the hump can be determined. A line
symmetrically intersecting the connection line of two
centers can be chosen as a threshold to separate two
qubit states. By adjusting the initial phase of the
readout pulse, the connection line of two hump centers
can be intentionally rotated to be horizontal in the
$IQ$ plane. Then the threshold is the voltage $I$ of
the midpoint between two centers. The threshold $I$
value has been shifted to 0 in the shown figure.

The number of data points is integrated with respect
to the $Q$ axis and the histogram of $I$ is shown in
Fig.~\ref{fig_n01}(c). For the ground state $|0\rangle$,
the main Gaussian hump is on the right side with
$I$ larger than $0$. For the excited state $|1\rangle$,
the main Gaussian hump is on the left side with $I$
smaller than $0$. For both $|0\rangle$ and $|1\rangle$,
there are still scattered data points on the other side
of the threshold line. The readout fidelity of the
ground (excited) state is defined as the fraction of
counted points with $I$ larger (smaller) than $0$ over
$C_\mathrm{tot}$, leading to $F_\mathrm{qB}^{0}\approx97.3\%$
and $F_\mathrm{qB}^{1}\approx90.3\%$ in Fig.~\ref{fig_n01}c.
For the other qubit $Q_\mathrm{A}$, the two readout
fidelities are $F_\mathrm{qA}^{0}\approx96.1\%$ and
$F_\mathrm{qA}^{1}\approx93.1\%$. The readout fidelity
for the excited state is normally smaller than that
for the ground state, mainly due to the qubit decay
error in the measurement process.

The BD1 board enables a multiplexed dispersive readout,
in which readout pulses at different frequencies can be
multiplexed in the same readout line~\cite{chenyuAPL12}.
The readout signal is finally demodulated to different channels,
giving the qubit state result for each individual qubit.
In the current version, the FPGA algorithm in BD1 admits
a simultaneous measurement of eight qubits.

In the feedback control, a signal qubit is chosen to
provide a feedback tag for the superconducting qubit system.
By comparing each measured $I$ with the threshold value,
the qubit state of the signal qubit is determined to be
at the ground or excited state. Correspondingly a feedback tag
can be generated and transferred from the feedback transmit
(TX) to the receive (RX) channels of other FPGA boards (called
control board). After receiving a feedback tag, the control
boards (BD2 and BD3) can output a conditional shaped pulse.
The DAC output is up-converted by the $IQ$ mixer and
another microwave source (LO2 with $f=4.800$ GHz) to the
qubit transition frequency. The different feedback-tag-controlled
pulses are transferred through $XY$ control lines to the
corresponding qubits to implement a conditional quantum gate.
For example, if Q$_\mathrm{A}$ is designated as the signal qubit,
the feedback-tag-controlled output of BD2 is a feedback control
on Q$_\mathrm{A}$, and the similar output of BD3 is a
feed-forward control on Q$_\mathrm{B}$. The DAC output of
BD2 and BD3 also provide pulses for general qubit operations
of Q$_\mathrm{A}$ and Q$_\mathrm{B}$, respectively. With the
integrated multi-boards and ISA (see Appendix D), this scheme can be
extended to a superconducting system with many qubits,
and multiple signal qubits can be chosen for the feedback
control in each subgroup of qubits.

\begin{figure}[htp]
\centering
\includegraphics[width=0.8\columnwidth]{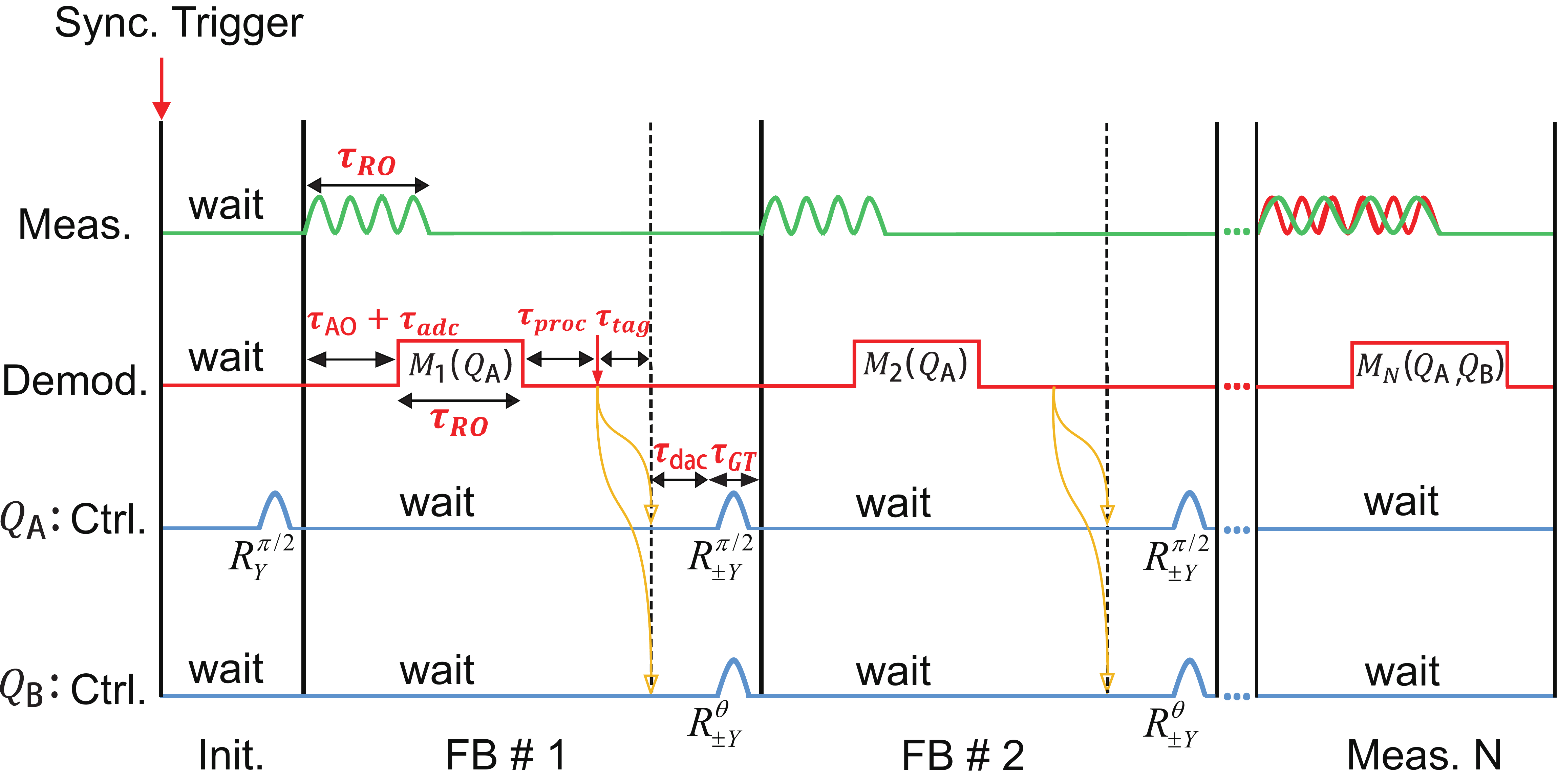}
\caption{Schematic diagram of the pulse sequence. From top to bottom are readout pulses,
demodulation windows and control pulses of the signal qubit Q$_\mathrm{A}$ and the target qubit Q$_\mathrm{B}$.
The sequence is synchronized by a trigger signal.
}
\label{fig_n02}
\end{figure}

The three boards and the microwave source are phase-locked
to an external 10 MHz clock. The timing between different
boards are synchronized, with the time dependent pulse sequence
shown in the schematic diagram in Fig.~\ref{fig_n02}. The
feedback latency in the feedback loop can be correspondingly
determined. In all related time scales, $\tau_\mathrm{RO}$
is the readout pulse length and $\tau_\mathrm{GT}$ is the
pulse length of the conditional gate. These two scales are
variable parameters in the feedback loop. Starting from the
readout pulse, there is a total delay in the low-temperature
analog devices and cables, $\tau_\mathrm{AO}$, before the
signal returns to the ADC. In Appendix C, we carefully explain
other latencies of room-temperature electronics. The
latency $\tau_\mathrm{adc}$ is introduced by the ADC,
which is about $16$ ns. The demodulation processing takes
a delay of $\tau_\mathrm{proc}=32$ ns, determined by
the clock period (4 ns) and number of flip-flops used
in the FPGA fabric. The latency $\tau_\mathrm{tag}$ is
measured to be 24 ns on the oscilloscope. After receiving
the feedback tag, there is another delay $\tau_\mathrm{dac}$
(68 ns) before the conditional gate pulse is sent out.
We define a total delay for the room temperature electronics, $\tau_\mathrm{tot}=\tau_\mathrm{adc}+\tau_\mathrm{proc}+\tau_\mathrm{tag}+\tau_\mathrm{dac}$,
which is 140 ns in the current status. The optimized
$\tau_\mathrm{tot}$ enables a fast feedback control
in our setup.

\begin{figure}[t]
\centering
\includegraphics[width=0.6\columnwidth]{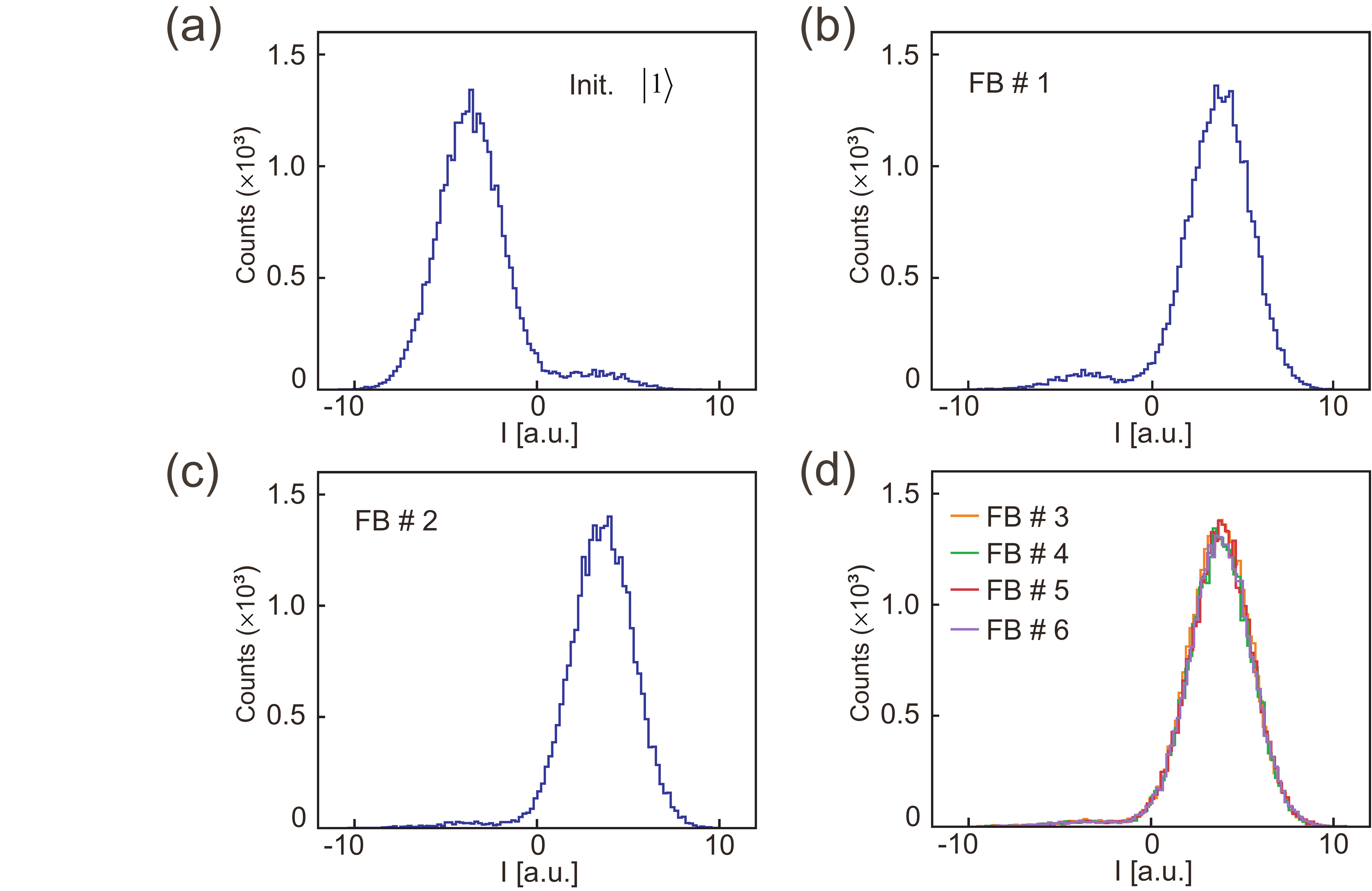}
\caption{(a) The histogram of $I$ signal for qubit Q$_\mathrm{A}$ initialized at the excited state $|1\rangle$.
(b) The histogram of $I$ signal when qubit Q$_\mathrm{A}$ is reset to the ground state $|0\rangle$ by a feedback
control.
(c) The histogram of $I$ signal when qubit Q$_\mathrm{A}$ is further reset to $|0\rangle$ by the 2nd feedback control.
(d) The histogram of $I$ signal when qubit Q$_\mathrm{A}$ is consecutively reset to $|0\rangle$ by multiple
feedback controls.
}
\label{fig_n03}
\end{figure}

\section{Results}
\label{sec3}
The feedback control is first applied to reset a qubit to the
ground state $|0\rangle$, which can simply prove the realization
of a feedback function~\cite{RistePRL12}. The Xmon qubit
Q$_\mathrm{A}$ is utilized for this single-qubit experiment.
The idle qubit is initially prepared at the excited state
$|1\rangle$ with a calibrated $\pi$ pulse. Afterwards a
feedback-reset-gate is applied, which includes a dispersive
qubit state measurement and a conditional control pulse.
If the qubit is measured to be at the excited state, the
conditional control is designated as a $\pi$ pulse to reset the
qubit to the ground state. If the qubit is measured to be at the
ground state, the control gate is instead designated as an
empty sequence but waiting for the same time as
the duration of the $\pi$ pulse. Afterwards the qubit is
measured again to check the effectiveness of the feedback-reset-gate.
The same procedure has been repeated for $C_\mathrm{tot}=3\times10^{4}$ times.
We integrate the number of data points with respect to the $Q$ axis and
show the histograms of $I$ in Fig.~\ref{fig_n03}. In
Fig.~\ref{fig_n03}(a), the main distribution of data points shows a
hump with Gaussian distribution on the left side, with $I$ smaller
than $0$. The calculated ratio of the excited-state probability
is $93.0\%$. For the qubit-state measurement after the conditional
feedback control, most of the data points in Fig.~\ref{fig_n03}(b)
concentrate on the right side, also showing a hump with Gaussian
distribution. The main distribution of qubit state shifts from the
left side to the right side means that the initialized qubit
(at excited state $|1\rangle$) is reset to the ground state $|0\rangle$,
by the conditional feedback control. In Fig.~\ref{fig_n03}(b),
a small hump can also be observed on the left side with a
probability of $6.3\%$. Compared to the initialized ground state, this
hump is a little bit higher~\cite{RistePRL12}.
The feedback function can be enabled consecutively for multiple
times~\cite{HuNat19,AndersenNpjQi19,NegnevitskyNat19}.
After the 2nd measurement, we apply another conditional gate based
on the measurement result. Then a 3rd measurement is taken to
check the result, with the histogram of $I$ shown in
Fig.~\ref{fig_n03}(c). Compared with Fig.~\ref{fig_n03}(b),
Fig.~\ref{fig_n03}(c) shows a similar distribution, but with
the small hump on the left side depressed to a probability of $3.5\%$~\cite{RistePRL12}.
The similar feedback control has been repeated for six times,
with the multiple measurements shown in Fig.~\ref{fig_n03}(d). The consecutive
feedback controls lead to a steady state. For the result with six
feedback loops, the calculated ground state probability is $96.5\%$.

\begin{figure}[t]
\centering
\includegraphics[width=0.6\columnwidth]{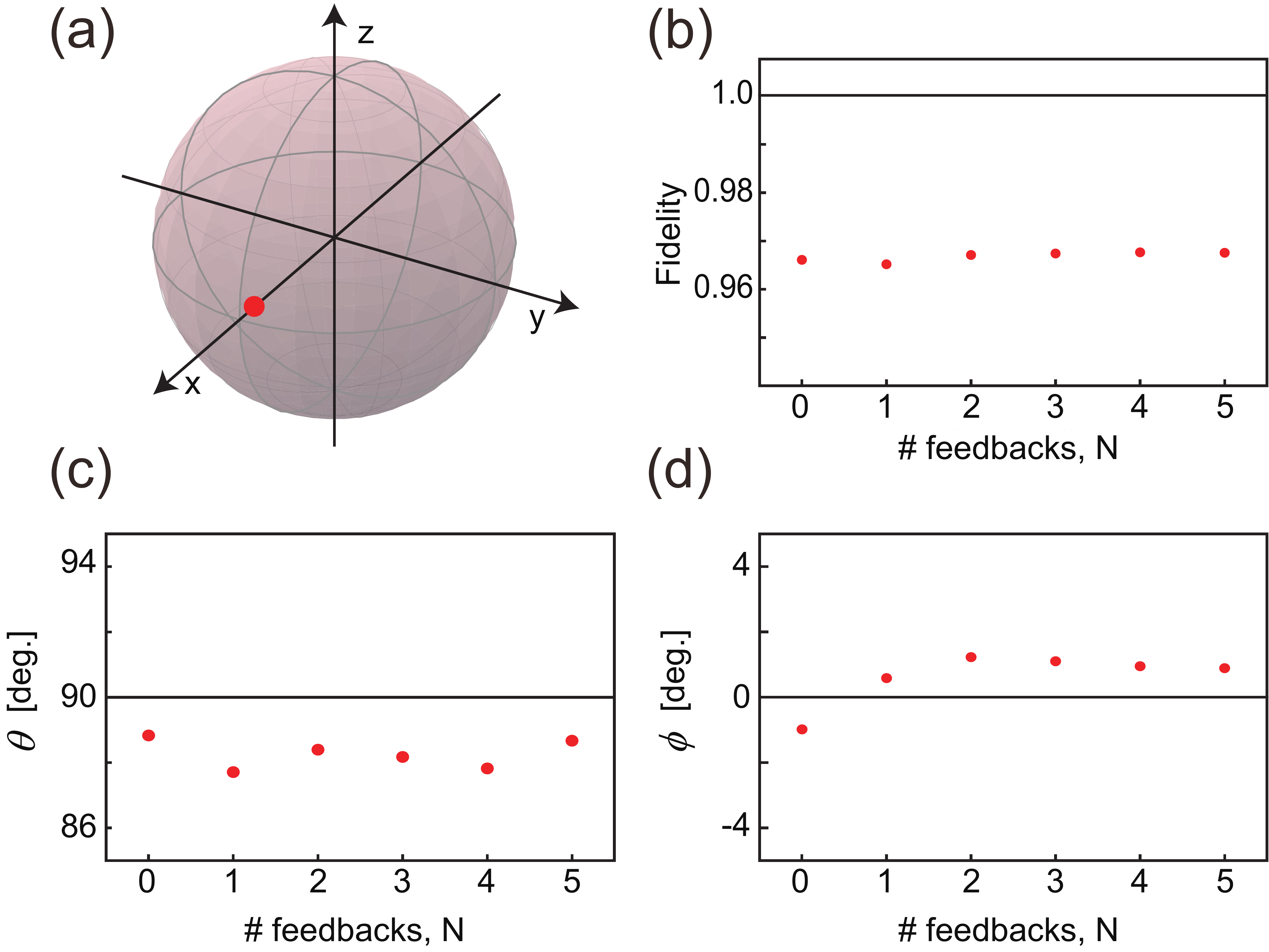}
\caption{
(a) After initialized at the superposition state $(|0\rangle+|1\rangle)/\sqrt{2}$, the qubit Q$_\mathrm{A}$ is
measured with a quantum state tomography (QST), with the measured Bloch vector displayed as a single point on the
Bloch sphere.
(b) With qubit Q$_\mathrm{A}$ consecutively frozen at the superposition state $(|0\rangle+|1\rangle)/\sqrt{2}$ by multiple
feedback controls, the qubit state are measured by the QST and the state fidelity are correspondingly calculated
and displayed.
(c) From the QST measurements, the polar angle $\theta$ of the Bloch vector are extracted and displayed.
(d) The azimuth angle $\phi$ of the Bloch vector are extracted and displayed.
}
\label{fig_n04}
\end{figure}

Reset the qubit to the ground state is a special application of
the feedback control. Because of the flexibility of the conditional
gate, the feedback control can be applied to prepare the qubit
to any quantum state on the Bloch sphere~\cite{CampagnePRX13}.
The superposition state $(|0\rangle+|1\rangle)/\sqrt{2}$ is
taken as an example. The qubit is initialized at
$\Psi=(|0\rangle+|1\rangle)/\sqrt{2}$ by a $\pi/2$ rotation
around the $y$ axis. Then the qubit is measured with a projective
measurement. If the qubit is projected to the ground (excited)
state, a conditional $\pi/2$ pulse around the $y$ ($-y$) axis
encoded in the algorithm is applied in the feedback control.
No matter what the intermediate measurement result is, the qubit
is brought back to the state $(|0\rangle+|1\rangle)/\sqrt{2}$.

A quantum state tomography (QST) is applied to measure the state
after the feedback control (see Appendix A)~\cite{ChuangBook,SteffenPRL06}.
For any qubit state, a density matrix can be expanded as
$\boldsymbol{\rho}=1/2(\boldsymbol{I}+\boldsymbol{r}\cdot\boldsymbol{\sigma})$,
where $\boldsymbol{I}$ and $\boldsymbol{\sigma}$ are
the identity and pauli matrices. Determined with the QST measurement,
the Bloch vector $\boldsymbol{r}=(x,y,z)$ can be depicted
as a single point on the Bloch sphere. In this experiment, the same
procedure is repeated for $C_\mathrm{tot}=1.5\times10^{4}$ times
for every QST projection direction.
As shown in Fig.~\ref{fig_n04}(a), the Bloch vector of the
prepared state is plotted, which points
to the $x$ axis and is consistent with the expected state of
$(|0\rangle+|1\rangle)/\sqrt{2}$. With $\Psi$ the known ideal state,
the fidelity of the prepared state $F=\langle\Psi|\boldsymbol{\rho}|\Psi\rangle$
is calculated to be $96.6\%$. Similar to the reset experiment, the qubit
state can be prepared to the designated state by consecutive
feedback controls. All the intermediate qubit-state readout
is a projective measurement. After a finite times of conditional
feedback controls, the final qubit state is measured with a QST.
For consecutive states, $\theta$ and $\phi$ of the
Bloch vector $\boldsymbol{r}$ are displayed in Fig.~\ref{fig_n04}(c)
and \ref{fig_n04}(d), respectively. The error of $\theta$ is
all smaller than $2.5^\circ$ and the error of $\phi$ is all
smaller than $1.5^\circ$. As shown in Fig.~\ref{fig_n04}(b),
the fidelity of consecutive state is all larger than $96\%$.

\begin{figure}[t]
\centering
\includegraphics[width=0.6\columnwidth]{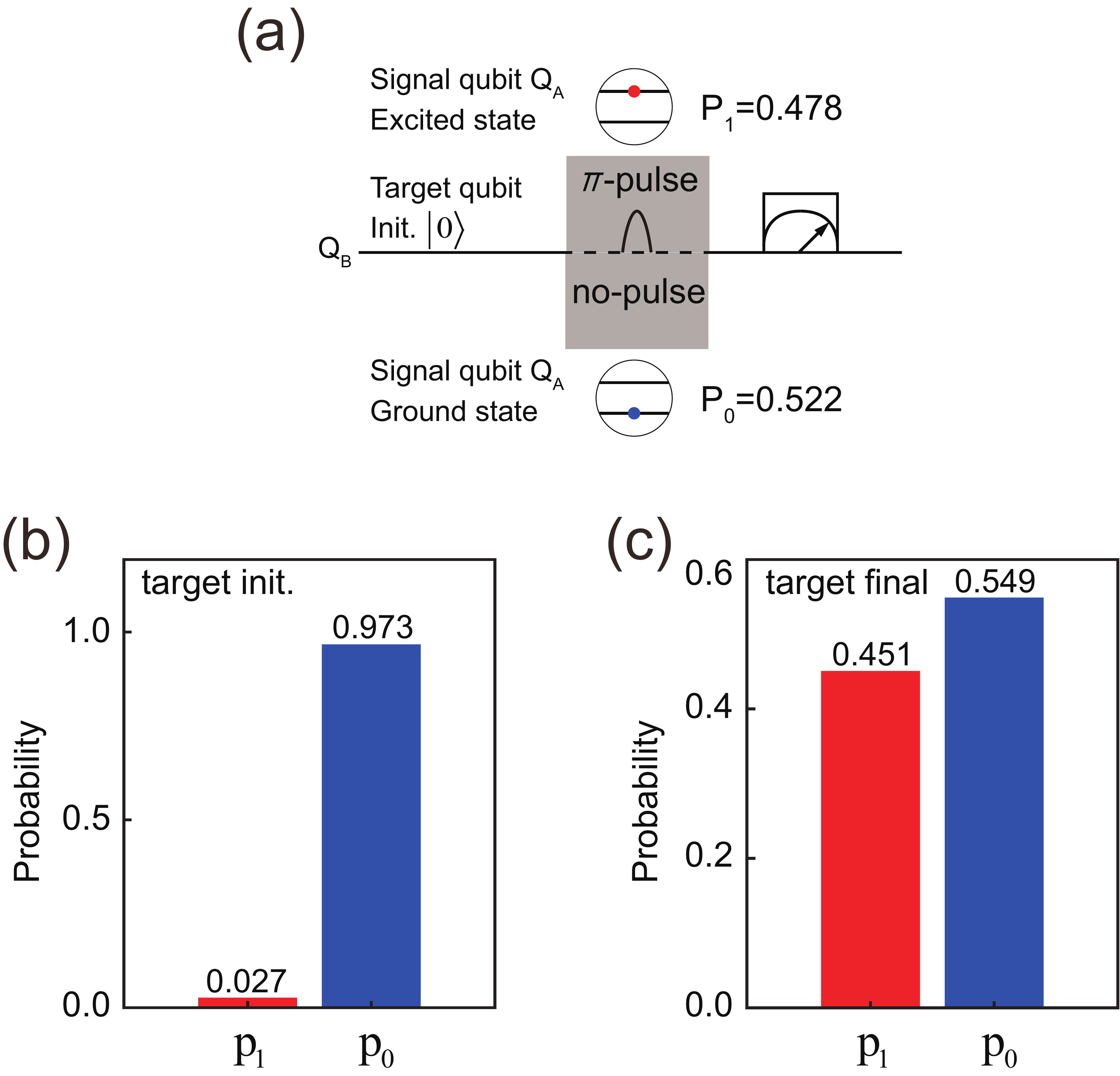}
\caption{
(a) Schematic diagram showing that the measurement of signal qubit is applied to
actuate a conditional pulse gate to the target qubit. With the signal qubit Q$_\mathrm{A}$
initialized at the superposition state $(|0\rangle+|1\rangle)/\sqrt{2}$, $P_0$ ($P_1$) is
the measured probability of $Q_\mathrm{A}$ for $I$ larger (smaller) than 0.
(b) With the target qubit Q$_\mathrm{B}$ initialized at the ground state $|0\rangle$, $p_0$ ($p_1$) is
the measured probability of $Q_\mathrm{B}$ for $I$ larger (smaller) than 0.
(c) The measured final probabilities of $Q_\mathrm{B}$ after the feed-forward control is applied.
}
\label{fig_n05}
\end{figure}

A feedback control applies a conditional gate to a qubit based
on the measurement result of the qubit itself. For a multi-qubit
system, a feed-forward control can provide special applications
in the quantum information processing~\cite{RisteNat13, SteffenNat13}.
To prove the feed-forward function, we choose $Q_\mathrm{A}$
as a signal qubit, and $Q_\mathrm{B}$ as a target qubit. The
feedback tag is generated based on the projective measurement
of the signal qubit. Then a feedback-tag-controlled conditional
pulse is issued to the target qubit. The two qubits are
physically separated by three qubits in the linear array. To
ignore any residue interaction, the signal qubit $Q_\mathrm{A}$
is biased away from the sweet point, at $\omega_\mathrm{qA}/2\pi=4.840$ GHz.
The frequency difference between two qubits is increased to
$|\omega_\mathrm{qA}-\omega_\mathrm{qB}|=240$ MHz, and two qubits
are effectively decoupled from each other. At this operation
point of $Q_\mathrm{A}$, the energy relaxation time $T_1$ is
9.3 $\mu$s and the pure dephasing time $T_2^*$ is 1.2 $\mu$s.

As a simple example, the signal qubit $Q_\mathrm{A}$ is initialized
at $(|0\rangle+|1\rangle)/\sqrt{2}$. The projective measurement
of the signal qubit is applied for the feed-froward control on
the target qubit $Q_\mathrm{B}$, as shown in the schematic diagram
in Fig.~\ref{fig_n05}(a). In this experiment, the same procedure
has been repeated for $C_\mathrm{tot}=1.5\times10^{4}$ times.
For the initialized $Q_\mathrm{A}$, there is a larger probability
of $P_0=52.2\%$ for the detection of $I$ signal larger than 0.
Correspondingly the probability of $I$ signal smaller than 0 is
$P_1=47.8\%$. For an ideal state of $(|0\rangle+|1\rangle)/\sqrt{2}$, there
should be an equal probability of $50\%$ for the qubit to
be projected to the state of $|0\rangle$ and $|1\rangle$. After
a measurement correction (see Appendix B), the corrected
populations of $Q_\mathrm{A}$ are $P_0^i=50.8\%$ and $P_1^i=49.2\%$,
close to the ideal equal probabilities. This result suggests that
the state preparation and measurement (SPAM) errors contribute to the
main error.


For the target qubit $Q_\mathrm{B}$, it is
initialized at the ground state $|0\rangle$, with a measured histogram
shown in the bottom left panel in Fig.~\ref{fig_n05}(b). For simplicity,
we integrate the number of data points in the $IQ$ plane for
$I$ both smaller and larger than zero, and show two ratio bars
for the ground and excited state probabilities. The probability
of $I$ larger than $0$ is $97.3\%$ for the initialized target
qubit $Q_\mathrm{B}$. For the feed-forward control, a $\pi$ (empty)
pulse is applied to $Q_\mathrm{B}$ if $Q_\mathrm{A}$ is measured to
be at the excited (ground) state. Afterwards the qubit state of $Q_\mathrm{B}$ is
measured for checking the function of the feed-forward control. From
an ensemble measurement of the target qubit $Q_\mathrm{B}$,
its state probability is determined as shown in the histogram
in Fig.~\ref{fig_n05}(c). With the signal
qubit in the state of $(|0\rangle+|1\rangle)/\sqrt{2}$, there is
approximately a $50\%$ probability
for the qubit to be pumped to the excited state. With a relative
smaller ratio ($P_1=47.8\%$) for $Q_\mathrm{A}$ to be in the excited
state, there is also a relative smaller ratio ($45.1\%$) for
$Q_\mathrm{B}$ to be pumped to the excited state.
The probability for $Q_\mathrm{B}$ to be at $|0\rangle$ ($|1\rangle$)
can be statistically calculated by $p^0_\mathrm{f}=p_0P_0+p_1P_1$
($p^1_\mathrm{f}=p_0P_1+p_1P_0$),
where $P_0/P_1$ and $p_0/p_1$ are the initialized probabilities of
$Q_\mathrm{A}$ and $Q_\mathrm{B}$, respectively. The calculated
result is $p^0_\mathrm{f}=52.1\%$ and $p^1_\mathrm{f}=47.9\%$.
After a measurement correction (Appendix A), the calculated result can
be calibrated to $p^0_\mathrm{f}=55.3\%$ and $p^1_\mathrm{f}=44.7\%$,
close to the experimental result of $p^0_\mathrm{f}=54.9\%$ and $p^1_\mathrm{f}=45.1\%$.

\begin{figure}[t]
\centering
\includegraphics[width=0.65\columnwidth]{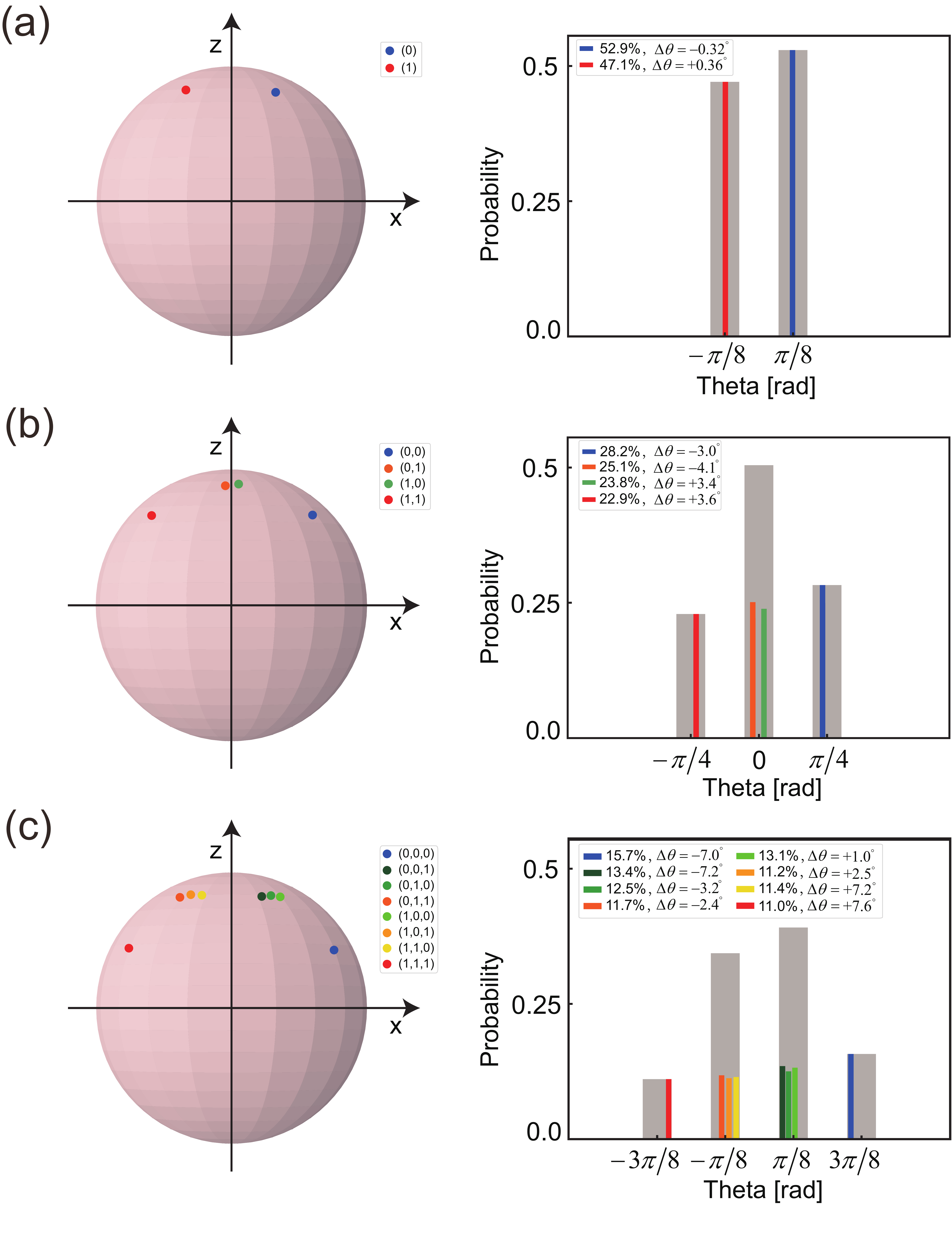}
\caption{
(a) QST result of the one-step random walk of the target qubit, which is drawn as data points of state vectors
on the $xz$-plane of the bloch-sphere. In the right panel, the polar angles of clockwise and anticlockwise
rotations of the random walk are shown by the blue and red bars, respectively. The grey bars are centered at ideal angles.
The height of both colored bars and grey bars is the percentage of collected data in corresponding QST measurement.
(b) QST result of the two-step random walk of the target qubit.
(c) QST result of the three-step random walk of the target qubit.
}
\label{fig_n06}
\end{figure}

For the random walk experiment, qubit $Q_\mathrm{A}$ is utilized
as a signal qubit and prepared at $(|0\rangle+|1\rangle)/\sqrt{2}$.
The projective measurement of this qubit produces a quantum
random number of 0 or 1, and a corresponding feedback tag to
control the target qubit. The target qubit $Q_\mathrm{B}$ is
initially prepared at the ground state $|0\rangle$
(the north pole in the Bloch sphere). Depending on the
1st measurement of the signal qubit, the feedback control
is a conditional pulse to rotate the target qubit away from the
north pole. For the signal qubit measured in the ground
or excited state, we rotate $Q_\mathrm{B}$ around the $y$-axis
or $-y$-axis with angle $\theta$, respectively.
After the operation of $R^{\theta}_y$ or $R^{\theta}_{-y}$, a collection
of tomography pulses are further applied for the QST measurement
of $Q_\mathrm{B}$. In the following experiment, the procedure is
repeated for $C_\mathrm{tot}=1.5\times10^{4}$ times for each
QST projection direction. Because we can record each measurement result
of the signal qubit, the measurement for the target qubit with
$R^{\theta}_y$ and $R^{\theta}_{-y}$ rotations can be separately
collected for the QST measurement. For an example of $\theta=\pi/8$,
the tomography result for the one-step random walk is shown in the
Bloch sphere in Fig.~\ref{fig_n06}(a). Looking at the $xz$-plane
from the $-y$-axis, the state vector is observed to rotate to
the right or left side of the north pole. In the right panel of
Fig.~\ref{fig_n06}(a), we show both measured angles of the
state vector in the QST. To distinguish the two random-walk directions,
the clockwise rotation is labeled with a positive angle while
the anticlockwise rotation is labeled with a negative angle.
The measured angles slightly deviate from the ideal value
with an error of $-0.32^\circ$ and $0.36^\circ$ for signal
qubit at $|0\rangle$ and $|1\rangle$, respectively.
The percentage of the collected data in QST is also shown for
the two vectors. There is a relative larger
ratio ($52.9\%$) for the target qubit to rotate clockwise, due
to the larger ratio for signal qubit at $|0\rangle$ (without correction).
The background grey bars are centered at ideal angles, while
the height of each grey bar is set to the corresponding
experimental result.

To apply consecutive feed-forward controls for the multi-step
random walk on the Bloch sphere, the signal qubit is simultaneously
frozen at $(|0\rangle+|1\rangle)/\sqrt{2}$ by a feedback
control at each step. After the 1st measurement of the signal
qubit, the target qubit is rotated clockwise or anticlockwise, with
an angle of $\theta=\pi/8$ or $\theta=-\pi/8$. After the 2nd
measurement of the signal qubit, the target qubit
is assigned to rotate again depending on the feedback tag,
no matter how the 1st-step rotation evolves. The target qubit is
then appended with tomography pulses for the final-state determination.
Here the feedback measurement can be collected to four groups
$\{00, 01, 10, 11\}$, with the two numbers representing the 1st and
2nd measurement results of the signal qubit. For each group of
ensemble measurement, the tomography result is collected for
extracting the final state on the Bloch sphere, as shown in
Fig.~\ref{fig_n06}(b). If a reversed walk is involved, the
qubit is rotated back to be close to the north pole, as shown
for the two vectors related with $\{01\}$ and $\{10\}$.
In the right panel of Fig.~\ref{fig_n06}(b), we present measured
angles of state vectors and corresponding percentages of collected
data for the two-step random walk. For signal qubit at state of
$\{00\}$ and $\{11\}$, the measured angles deviate from the ideal
value with an error of $-3.0^\circ$ and $3.6^\circ$, respectively.
For groups of $\{01\}$ and $\{10\}$, a small difference happens
between the percentages of two vectors, due to the
slight variation of probabilities for consecutive prepared state
of the signal qubit. The height of the central grey bar is set
to the addition of two percentages. A three-step random walk is
similarly actuated by the simultaneous feedback and
feed-forward control, with the results shown in
Fig.~\ref{fig_n06}(c). For groups of $\{000\}$ and $\{111\}$,
the measured angles deviate from ideal values with an error
of $-7.0^\circ$ and $7.6^\circ$, respectively. Compared with
angles related wtih $\{00\}$ and $\{11\}$ in the two-step random walk,
and angles related with $\{0\}$ and $\{1\}$ in the one-step random walk,
the angle error increases with the number of walk steps, which
is found to be related with the qubit decay and dephasing in the simulation
analysis (see Appendix E). For groups of $\{001\}$, $\{010\}$
and $\{010\}$, state vectors share similar vector angles
around $\theta=\pi/8$, and the height of the grey bar is
set to the addition of three percentages. The similar behavior
happens for groups of $\{101\}$, $\{011\}$ and $\{110\}$.
The measured random walk of target qubit on the Bloch sphere
proves both the realization of consecutive feedback/feed-forward
control and the precise quantum control in our superconducting
multi-qubit system.

\section{Summary}
\label{sec5}
We develop a fast FPGA-based feedback control system, with the latency of room-temperature
electronics optimized to 140 ns. The function of feedback control is proved by resetting the
qubit to the ground state and preparing the qubit to a designated superposition state with high-fidelity.
In two-qubit experiments, projective measurement of the signal qubit provides a feed-forward
control to the target qubit. The consecutive and simultaneous feedback and feed-forward control
enables a random walk of the target qubit on the Bloch sphere. Our experiment is a conceptually
simple benchmark for the feedback control, the realization of which proves that the control
system is appropriate for a multi-qubit experiment. Furthermore, the hardware and ISA can be
expanded to more complex feedback applications, such as the hardware accelerator of error
correction code and the compiler of high level quantum language.

\begin{acknowledgments}
The work reported here was supported by
the National Key Research and Development Program of China (Grant No. 2019YFA0308602, No. 2016YFA0301700),
the National Natural Science Foundation of China (Grants No. 11934010, No. 11775129),
the Fundamental Research Funds for the Central Universities in China, and the Anhui
Initiative in Quantum Information Technologies (Grant No. AHY080000). Y.Y. acknowledge
the funding support from Tencent Corporation. This work was partially conducted at the
University of Science and Technology of the China Center
for Micro- and Nanoscale Research and Fabrication.
\end{acknowledgments}

\appendix

\section{Quantum State Tomography (QST) Measurement}
\label{appendixA}
To fully determine the quantum state of a two-level qubit, we need to realize the quantum
state tomography (QST) measurement. The density matrix of either a pure or mixed state can be expanded as
$\rho = \frac{1}{2}\left(I + x \sigma_x + y \sigma_y + z \sigma_z\right)$,
with $I=|0\rangle\langle 0|+|1\rangle\langle 1|$. We introduce three Pauli operators,
$\sigma_x = |0\rangle\langle1|+|1\rangle\langle 0|$, $\sigma_y = -i|0\rangle\langle1|+i|1\rangle\langle 0|$,
and $\sigma_z = |0\rangle\langle0|-|1\rangle\langle 1|$, based on which a vector of Pauli
operators is represented by $\bm \sigma=(\sigma_x, \sigma_y, \sigma_z)$. The three projections, $x$, $y$ and $z$
along the three directions, determine a vector, $\bm r = (x, y, z)$, which is named as the Bloch vector.
The $z$-projection, $z=P_0-P_1$, is extracted from
a projective measurement of the qubit probability. To extract the $x$-projection, we rotate
the quantum state by an angle of $-\pi/2$ around the $y$-axis and the density matrix is changed to be
$\rho^\pr = U_y(-\pi/2) \rho U^+_y(-\pi/2)$,
where $U_\zeta(\theta)=\exp[-i\theta\sigma_\zeta/2]$ is an unitary operator for a rotation angle of $\theta$
around the $\zeta(=x, y, z)$-axis. Experimentally, the $U_\zeta(\theta)$ gate is realized by a calibrated
pulse with the frequency $\omega_{10}$.
The rotated density matrix is then given by $\rho^\pr=(I_2-z\sigma_x+y\sigma_y+x\sigma_z)/2$. The
population measurement determines the $x$-projection, $x=P^\pr_0-P^\pr_1$, where $P^\pr_0$ and $P^\pr_1$
are the populations of the ground and excited states in the rotated density matrix.
In practice, the $U_y(\pi/2)$ gate is also applied for the measurement
and the $x$-projection is averaged from the results under the operations of the
$U_y(\pi/2)$ and $U_y(-\pi/2)$ gates. The $y$-projection is similarly obtained using the operations
of the $U_x(\pi/2)$ and $U_x(-\pi/2)$ gates.

\section{Measurement Correction}
\label{appendixB}
From the readout fidelity measured for both ground ($F_0$) and excited ($F_1$) state, we could observe that
the measured population probabilities ($P^\mathrm{m}$) are often different from the ideal result ($P^\mathrm{i}$).
The relation between the ideal and measured populations can be expressed as
\be
\left(\ba{c} P_0^\mathrm{m}\\ P_1^\mathrm{m} \ea \right)=
\left(\ba{cc} F_0&{1-F_1}\\{1-F_0}&F_1 \ea\right) \left(\ba{c} P_0^\mathrm{i}\\P_1^\mathrm{i} \ea\right),
\label{eq_S1}
\ee
from which the ideal populations can be calibrated from measured population probabilities.

\section{Latency of Electronics}
\label{appendixC}
{\bf FPGA logic latency.} The digital circuit is designed on a
Xilinx Kintex-7 (XC7K325T) field-programmable-gate-array (FPGA).
We use SystemVerilog language to describe the digital circuit and use
the Xilinx Vivado Design Suite${\textregistered}$ to simulate,
synthesize and finally implement the design on FPGA~\cite{XilinxVDS}.
Data path logic for all the FPGA design runs at 250 MHz clock
frequency, leading to a 4 ns latency for each flip-flop. This flip-flop
latency is relatively shorter than that of other FPGA chips, such
as Xilinx Virtex-4 and Virtex-6~\cite{SalathPRAppl18,HuNat19}.

We include both signal detection and waveform generation in the multi-board
feedback control system. For the measure board, one FPGA chip is integrated
with two digital-to-analog-converters (DACs) and two
analog-to-digital-converters (ADCs). For the control board, one FPGA chip
is integrated with four digital-to-analog-converters (DACs).

\begin{figure}
\centering
\includegraphics[width=0.95\columnwidth]{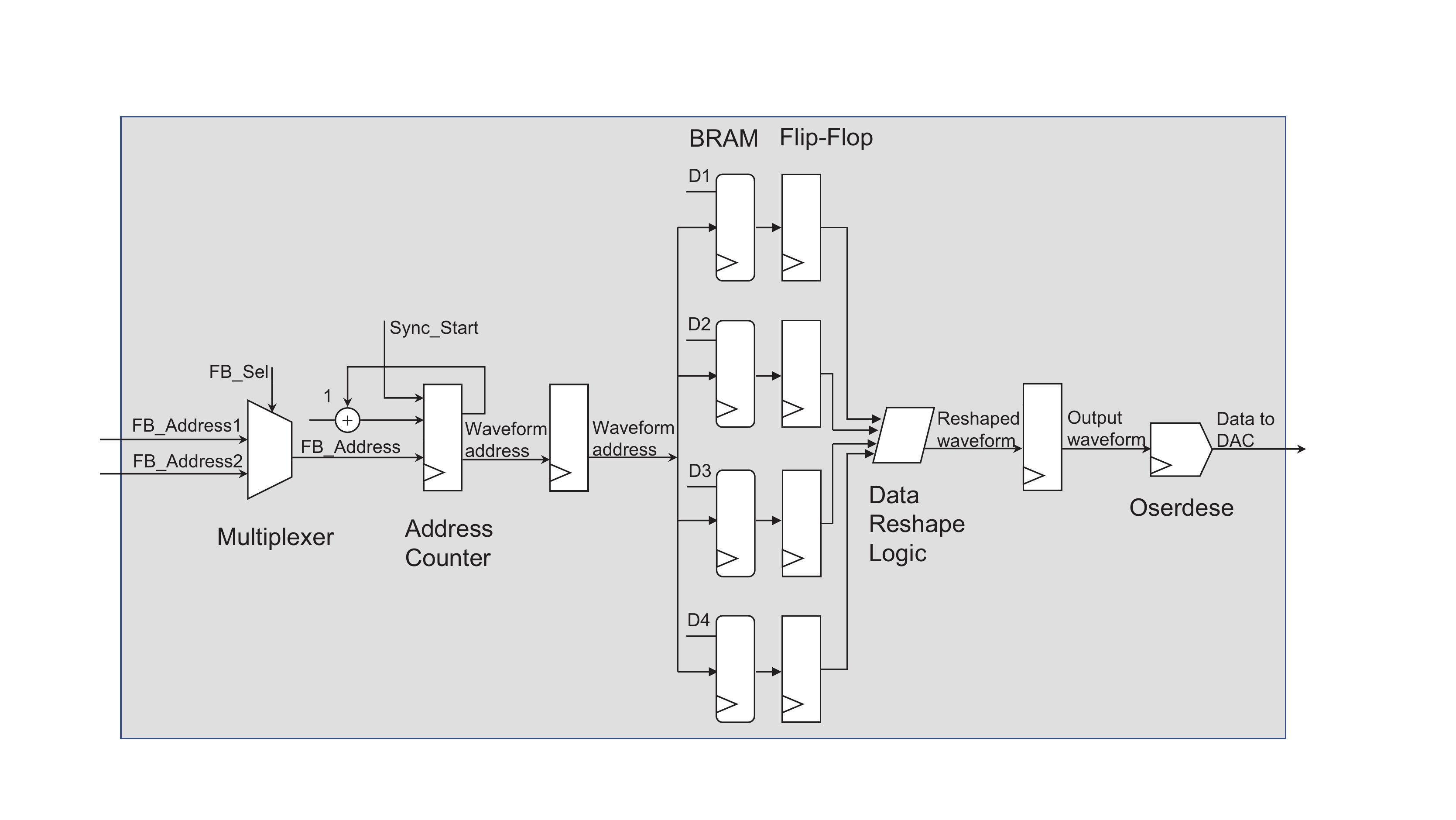}
\caption{A block diagram of data path for FPGA to send digital waveform to the DAC card.
}
\label{fig07}
\end{figure}

The DAC actuate latency is the time for FPGA to
send digital waveform to the DAC card after it receives the trigger signal.
A block diagram of the corresponding data path is shown in Fig.~\ref{fig07}.
The control pulse waveform is stored in Block Random Access
Memories (BRAMs) of the FPGA, the read address of which can be determined by a
feedback signal. One sample of digital waveform is readout from the BRAM
every clock cycle (4 ns) and then streamed to the DAC. We parallelize
BRAMs as 4 paths to achieve the total throughput of 1 GSPS, which
matches the sampling rate of DAC. Signal delay of the BRAM
output is a critical path, thus we insert a flip-flop after the output data of
BRAM to pipeline the data path. Because of the similar reason, we insert a flip-flop
both after the address counter and after the data reshape logic. With the
oserdes~\cite{XilinxIDDR} module introducing one more clock cycle,
the total DAC actuate latency is $6\times4=24$ ns.

\begin{figure}
\centering
\includegraphics[width=1.0\columnwidth]{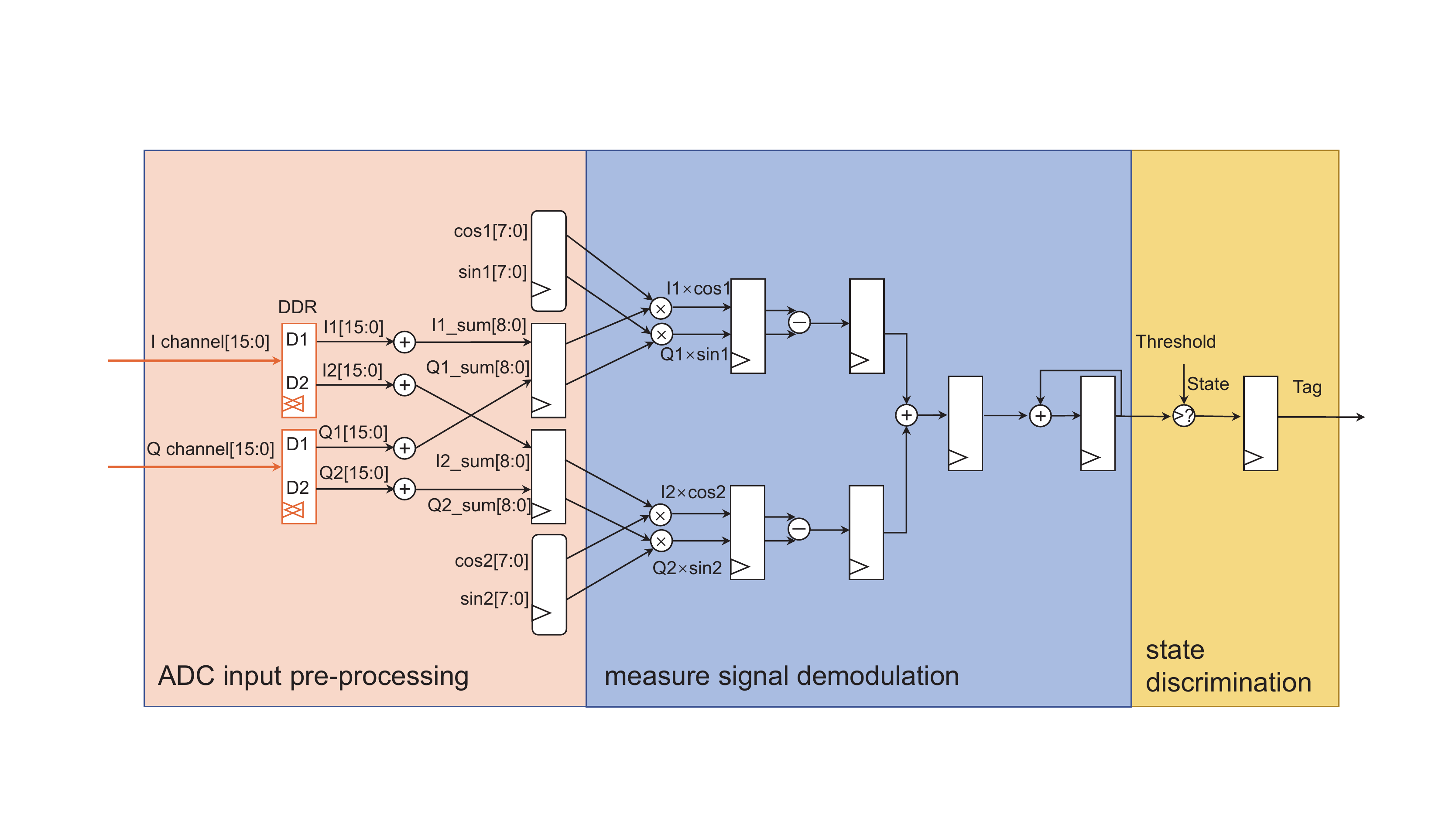}
\caption{A block diagram of data path for FPGA to measure signal modulation and make state discrimination.
}
\label{fig08}
\end{figure}

For the analog-to-digital-converter (ADC) signal processing latency, the block
diagram of the data path is shown in Fig.~\ref{fig08}. The path is composed
of three parts, the ADC input pre-processing, the measure signal demodulation,
and the state discrimination. In the first part, two channels of ADC output enter the FPGA.
Each ADC has a resolution of 8 bits, working at a conversion rate of 1 GSPS. Correspondingly
the ADC presents 2 adjacent digitized samples (16 bits) to the FPGA pad every 2 ns. With the
data path clocked every 4 ns, we use Xilinx¡¯s input-double-data-rate (IDDR)
primitive~\cite{XilinxIDDR} to register the ADC data twice every clock cycle. The DDR
module is configured as the SAME$\_$EDGE$\_$PIPELINED mode. In this mode,
it outputs pairs D1 and D2 to the FPGA logic at
the same clock edge, with a delay of 2 clock cycles relative to the
input~\cite{XilinxIDDR}. For every clock cycle, D1 is the first
2 ns samples data and D2 the second 2 ns samples data. To simplify the logic
design, 2 ns samples are adjacently summed up to a 9-bit data. Here we
halve the ADC sampling rate without losing much useful information, because
the ADC input signal is within the cutoff frequency of 250 MHz of the ADC
anti-aliasing filter. The summed 9-bit data is flip-floped before going to
the next stage.

The following stage of demodulation works in several independent and parallel channels. The
number of channels that can be implemented is limited by the available FPGA
resource, especially the DSP48E1 slice~\cite{XilinxDSP48E1}. In our current design,
we implement eight qubit channels, with each channel discriminating one
qubit state and generating one feedback tag. For the signal demodulation, the
digital signal $V_\mathrm{in}[t_n]=I_\mathrm{in}[t_n]+iQ_\mathrm{in}[t_n]$ is
multiplied with a reference signal, summed and accumulated for a window of
$N$ clock cycles for the final demodulation result. The complex output signal is

\be
\begin{aligned}
\tilde{V}_\mathrm{out}&=\tilde{I}_\mathrm{out}+i\tilde{Q}_\mathrm{out}\\
                      &=\sum_{n=1}^{N}\left(I_\mathrm{in}[t_n]+iQ_\mathrm{in}[t_n]\right)\times\left(\mathrm{cos}[\omega_rt_n]+i\mathrm{sin}[\omega_rt_n] \right)\\
                      &=\sum_{n=1}^{N}\left(I_\mathrm{in}[t_n]\mathrm{cos}[\omega_rt_n]-Q_\mathrm{in}[t_n]\mathrm{sin}[\omega_rt_n]\right)+i\left(I_\mathrm{in}[t_n]\mathrm{sin}[\omega_rt_n]+Q_\mathrm{in}[t_n]\mathrm{cos}[\omega_rt_n]\right)                   ,
\end{aligned}
\label{eq_01}
\ee
in which $\omega_r$ is the reference frequency. Each $\omega_r$ is determined by the readout
resonator of corresponding qubit.
In the demodulation processing, both the real and imaginary part are implemented,
with the final result $\tilde{V}_\mathrm{out}$ denoted as a single point
in the complex $IQ$ plane. The phase information of $\tilde{V}_\mathrm{out}$
can be adjusted by introducing an extra phase shift in
either the readout pulse or the reference signal.
Correspondingly the connection line of the qubit state hump centers
can be rotated to be horizontal in the $IQ$ plane. Then the real part
of the demodulation result can be used to discriminate the qubit state
and generate a feedback tag. For simplicity, the block diagram in
Fig.~\ref{fig08} only shows the real part processing. Four flip-flops
introduce 4 clock cycles of latency in the demodulation stage. In the third part,
the accumulated $I_\mathrm{out}$ component is compared to the feedback threshold and a feedback
tag is generated, with one extra flip-flop. The total ADC signal processing latency is
then $\tau_\mathrm{proc}=8\times 4=32$ ns. Note that the threshold is pre-loaded to
the FPGA from the host computer, and eight different thresholds can be set for
different qubit channels.

We also apply a waveform record module for timing analysis. The input signal
is similarly through the DDR as in the ADC signal demodulation logic. Afterwards
the record function is triggered by the synchronous start signal. The ADC input
signal is written to the SRAM after 3 clock cycles, with a latency of 12 ns.
Using this module, we can experimentally measure the delay of the qubit
readout chain. When including the wiring in the DR, the total chain delay
is 160 ns. Bypassing the long microwave coax lines in the dilution refrigerator,
the total chain delay or the homemade electronics delay is measured to be 96 ns.

{\bf Inter board communication latency.} In the multi-qubit feedback control
system, the measure board generates feedback tags, 1 or 0, based on the
discriminated qubit state. In our current design, tags from 8 parallel channels are
encapsulated to a packet and transmitted to other control boards (BD2, BD3) through
the tag transmission logic. Ethernet cable is used as the physical channel between the
measure board and the control board. The cable includes both the forward
and receive lines, and tags are sent in the direction from the measure board
to the control board. The channel width is thus 2 in one ethernet cable,
meaning a two-bit data can be transferred simultaneously. The feedback tag
packet is designed as the following format,

\begin{table}[ht]
\begin{tabular}{|c|c|c|c|c|}
\hline
~~~~~~1~~~~~~~& ~~~Qubit1~~~~& ~~~Qubit3~~~~&~~~Qubit5~~~~&~~~Qubit7~~~~\\
\hline
~~~~~~1~~~~~~~& ~~~Qubit2~~~~& ~~~Qubit4~~~~&~~~Qubit6~~~~&~~~Qubit8~~~~\\
\hline
\end{tabular}
\end{table}

\noindent in which two rows represent parallel lines of the two-bit data. The first
column of 2 bits is the head of the tag packet, indicating the beginning of the
tag communication. The two-bit head of packet brings one clock
cycle before the control board receives the tag. The following bits include tag
results of different qubits. If only some of the eight channels are utilized,
a default tag value of 1 is set for all unused channels.

\begin{figure}
\centering
\includegraphics[width=0.6\columnwidth]{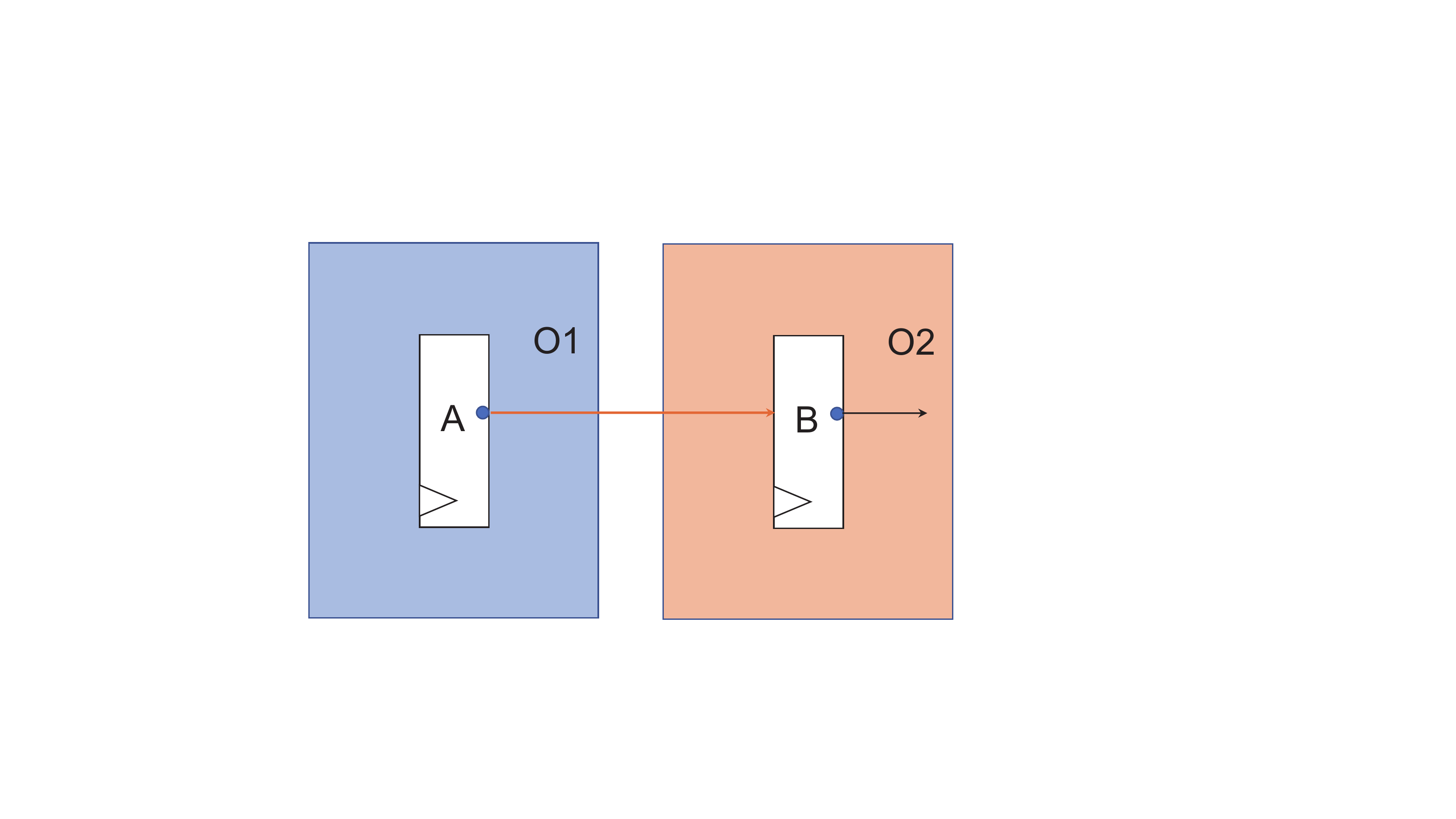}
\caption{A block diagram of data path for the inter board communication.
}
\label{fig09}
\end{figure}

Experimentally, we choose a cable whose length is 1.5 m. With a propagation
of 0.6 times the speed of light, the signal delay in the ethernet cable
is 8.3 ns. We also estimate a delay of about 8 ns on the PCB trace and the
pad to flip-flop delay inside of the FPGA. In the FPGA design, we route the
tag output of the measure board (the output of flip-flop A in Fig.~\ref{fig09})
to an output pin of the FPGA (labeled O1). We also route the flip-flopped
tag input (the output of flip-flop B in Fig.~\ref{fig09}) to its FPGA's
output pin (labeled O2). We probe the two signals at O1 and O2 using an
oscilloscope to measure the delay of a tag packet, which
is around 20 ns and close to the wiring delay plus one flip-flop delay.
Including an extra cycle of packet head, the total latency of feedback
tag communication is then $\tau_\mathrm{tag}=$24 ns.

{\bf Digital/Analog conversion latency.} The digital to analog conversion
latency is defined as the interval between FPGA sending out the digital waveform
data and the DAC daughter card outputting the base-band waveform to the $IQ$
mixer. Similarly, the analog to digital conversion latency is the interval
between ADC daughter card receiving the readout waveform and the FPGA receiving
the digitized waveform data. Both the DAC and ADC latency are mainly determined by the
DAC and ADC chip we selected.

We measure the delay of $\tau_\mathrm{dac}=68$ ns with an oscilloscope. It is
the time elapsed from the start trigger of the board to the first point
of the signal at the input of its $IQ$ mixer. Three parts contribute to the DAC latency
$\tau_\mathrm{dac}$. One is the internal latency of the DAC chip. From the
data sheet of AD9736 chip~\cite{DAC}, the analog output changes ($35+4=39$)
clock cycles (1 ns/cycle) after the input data changes, in which the 4 clock
cycles is the internal first-in-first-out (FIFO) latency we set.
The second is the DAC actuate latency 24 ns, coming from the total pipeline stages
of our customized FPGA logic as shown in Fig.~\ref{fig07}. The rest latency
is $68-39-24=5$ ns, which is expected due to the off-chip PCB trace and
other small components.

The ADC pipeline delay is the number of clock cycles needed for analog
to digital conversion. In the datasheet of ADC081000 chip~\cite{ADC},
this delay is 8 ns if working at the sampling rate of 1 GSPS. There is
another 2.7 ns delay for digital data going out of the ADC chip~\cite{ADC}.
Experimentally, we actuate a square pulse and trigger the measure board
to record it to the BRAM in FPGA, using the waveform record module of
the measure board. The timing lag of the recorded square pulse is measured
to be 96 ns, which indicates the base-band loop back time. This lag is
composed of $\tau_\mathrm{dac}$, measure pulse recording latency (12 ns),
the internal latency of the ADC chip, and the off-chip signal delay of
the ADC card. We can estimate the off-chip signal
delay as $96-68-12-8-2.7 = 5.3$ ns. The ADC latency is correspondingly
estimated to be $\tau_\mathrm{adc} = 8+2.7+5.3=16$ ns. The total feedback
latency of room-temperature electronics is then
$\tau_\mathrm{tot}=\tau_\mathrm{adc}+\tau_\mathrm{proc}+\tau_\mathrm{tag}+\tau_\mathrm{dac}=140$ ns.
In the future the latency can be further improved by dealing with
different parts, such as choosing DAC/ADC chips with better specifications.
We can also choose system-on-chip (SoC) that further integrate
hardware components like the FPGA, DAC, ADC, and filter on a single chip.

\section{Board Programming Model}
\label{appendixD}
Board programming model is also critical in the feedback control system. In Ref.~\cite{FuMICRO17},
the authors combine the codeword-triggered pulse generation and queue-based event timing control
to a centralized quantum control box. Similarly, C. A. Ryan \emph{et al.} use the super-scalar architecture and dispatch instructions to different working engines~\cite{RyanRSI17}. In our work
we adopt a different strategy and define a measure/control instructions set
architecture (ISA) for the deployment of quantum feedback tasks.

This ISA includes multiple instructions executed by multiple boards, and
it directly handles the pulse streaming and digital signal processing in measurement.
For a new experiment, digital pulse waveforms and all instructions
are loaded to each board from the host computer. The host also loads
control information to the registers of each FPGA, such as, 8-qubit channels
threshold of measure board, the tag select number of control board.
Each board starts to execute its instructions after receiving the synchronous trigger signal.

A measure instruction contains the information for the qubit demodulation as below.
\begin{table}[ht]
\begin{tabular}{|c|c|c|c|}
\hline
~Channel mask[7:0]~~& ~Repetition[3:0]~~& ~Delay[15:0]~~& ~Length[7:0]~~\\
\hline
\end{tabular}
\end{table}

\noindent Channel mask is 8 bits and each bit enables a qubit demodulation channel.
Repetition is the number of consecutive feedbacks. Delay is the number of cycles
delayed before the demodulation (see Fig. 2). Length is the
number of cycles of the demodulation window.

A control instruction for one board contains the information to generate the control pulse.
\begin{table}[ht]
\begin{tabular}{|c|c|c|c|c|c|}
\hline
~~Opcode[3:0]~~& ~~Index0[7:0]~~&~~Index1[7:0]~~&~~Address0[19:0]~~& ~~Address1[19:0]~~\\
\hline
\end{tabular}
\end{table}

\noindent Opcode directs the type of instructions. If $\mathrm{op}=0$,
the board will terminate the execution of the instruction. If $\mathrm{op}=1$,
when the current item is completed, the board will execute the next instruction.
If $\mathrm{op}=2$, when the current item is completed, the board will jump to
the instruction at index0. If $\mathrm{op}=3$, when the current item is completed,
the board will jump to the instruction conditioned on the received feedback tag.
The tag for the next instruction must be valid before the current instruction finishes.
Index0 and index1 are the address of next instruction, conditioned on the tag.
Address0 and address1 determine the segment of waveform in the BRAM, with the digital
waveform sequence starts from address0 and ends at address1.

This ISA is reduced to basic functions of reading the pulse waveform and
demodulation reference from the main memory, which brings flexibilities
and several advantages. First, because the timing scheduling is set by the
host computer, we can interleave instructions for fine-grained gate timing
optimization in feedback. In contrast, the centralized superscalar architecture
suffers great timing cost in the feedback trigger synchronization (210 ns) and
address jumping (53 ns)~\cite{RyanRSI17}. Second, with the pulse waveform
and measure reference pre-loaded to each board, we can reuse the memory
by only loading new gates for the next task, which greatly reduces the
host-board communication. Third, runtime variables can be set on the fly
by the host computer. This feature can be very useful when adaptively
calibrating the multi-qubit system, e.g., the adjust of feedback threshold.
In general, the ISA naturally supports the feedback/feed-forward control
with multiple signal qubit and multiple target qubit. On the other hand, this
ISA mimics features of a traditional computer model, like memory operation,
branching, MIMD (multiple instruction, multiple data), therefore it is
promising to integrate the ISA with classical processors. In the future,
more complex functions can be built upon this ISA, such as the compiler
of quantum feedback program, hardware accelerator of the error correction
code and quantum feedback runtime.

\section{Simulation of the random walk result}
\label{appendixE}

We numerically simulate the target qubit's random walk result, using the master equation,

\be
\begin{aligned}
\dot{\rho}(t)=-\frac{i}{\hbar}[H(t),\rho(t)]+\sum_n \frac{1}{2}[2C_n\rho(t)C_n^{+}-\rho(t)C_n^{+}C_n-C_n^{+}C_n\rho(t)],
\end{aligned}
\label{eq_a01}
\ee
in which $C_n=\sqrt{\gamma_n}A_n$ are coupling terms through which the
system couples to the environment. $A_n$ are the Lindblad operators
and $\gamma_n$ are the corresponding rates. For Xmon-type qubits,
system-environment coupling is related with two independent channels,
characterized by the relaxation term $A_1=|0\rangle\langle1|$ and pure
dephasing term $A_2=|1\rangle\langle1|$, with $\gamma_1=\frac{1}{T_1}$
and $\gamma_2=\frac{2}{T_2^*}$, respectively.

\begin{figure}
\centering
\includegraphics[width=1.0\columnwidth]{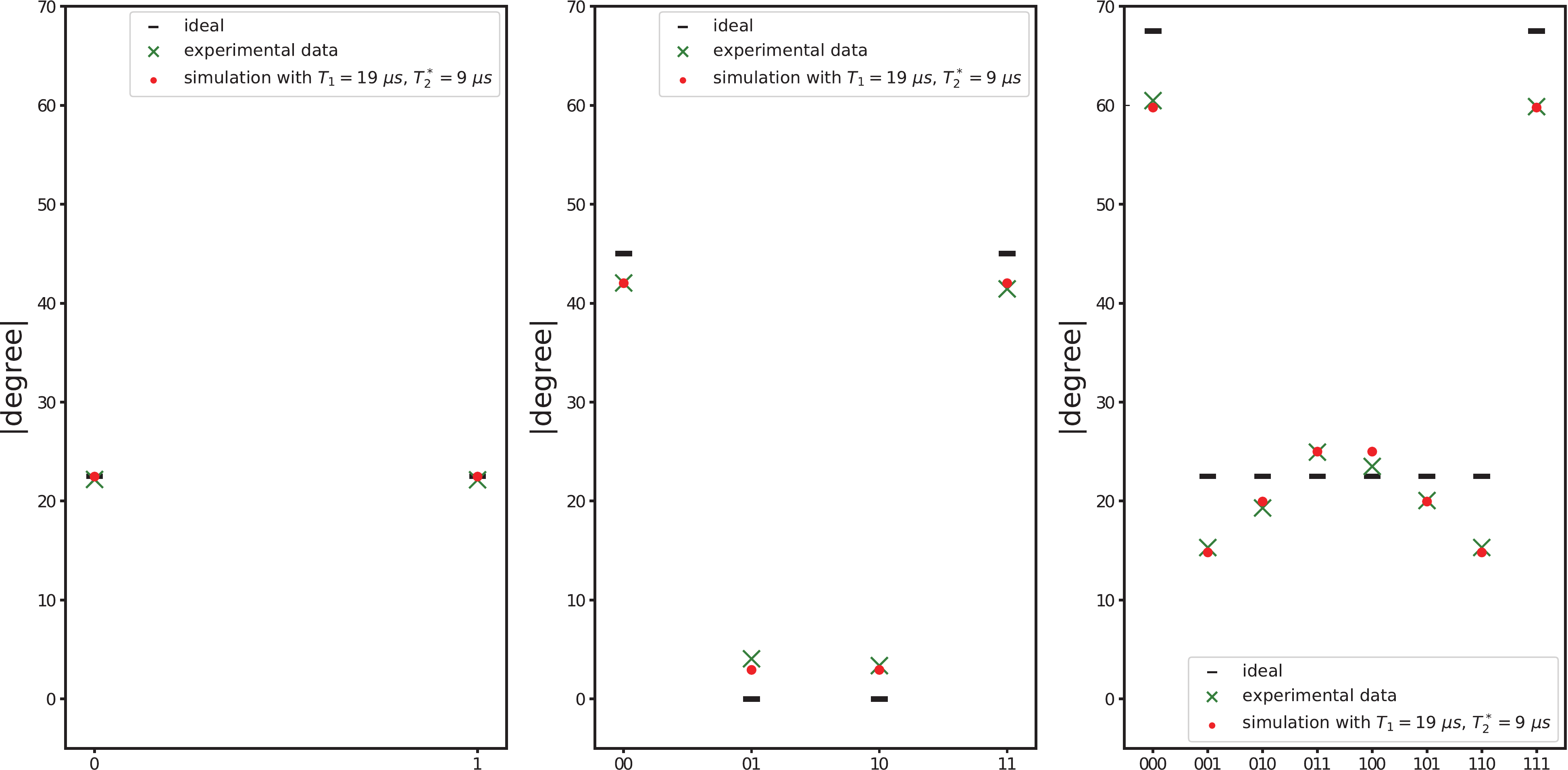}
\caption{Comparison between the experimental data and numerical simulation
of target qubit rotation angles. From left to right,
the three panels represent results of the one-step, two-steps and three-steps
random walk process. The horizontal axis labels different groups classified by
the measurement of signal qubit.
}
\label{fig10}
\end{figure}

When we simulate the random walk with infinite $T_1$ and $T_2^*$,
the rotation angle $\theta$ of the target qubit is almost the same as
the ideal values. When parameters are set as $T_1=19$ $\mu$S and
$T_2^*=33$ $\mu$S, the simulation result reproduces a deviation
of rotation angle of the target qubit away from ideal values (not shown),
with the same tendency as that in the experimental data. However,
a remaining difference exists between the experimental and simulation
result. Here we mention a technical
detail in the experimental process. During the random walk process
of the target qubit, the measurement of signal qubit is found to
induce an ac-Stark shift of the target
qubit~\cite{LiuPRX16,AndersenNpjQi19,SchusterPRL05}, represented by
a small detuning of $\sim0.05$ MHz (or 15 degree phase shift per step).
A square-shaped $Z$-pulse is applied to the target qubit to
compensate this phase offset before the next step. We suspect that
the signal qubit's readout pulse may account for a larger pure
dephasing rate of the target qubit. After adjusting the parameter
$T_2^*$, we found the simulation with $T_1=19$ $\mu$S and
$T_2^*=9$ $\mu$S is well consistent with the experimental data, as
shown in Fig.~\ref{fig10}. This simulation result indicates that the
qubit decay and dephasing induce the deviation of the roation angle
away from ideal values.

\end{document}